\documentclass[aps,prl,twocolumn,showpacs,superscriptaddress,groupedaddress]{revtex4-1}  
\usepackage{graphicx}  
\usepackage{dcolumn}   
\usepackage{bm}        
\usepackage{amssymb}   

\begin{document}



\title{Nuclear motion is classical}
\author{Irmgard Frank} 
\affiliation{Universit{\"a}t Hannover}
\date{\today}

\begin{abstract}
The notion from ab-initio molecular dynamics simulations that nuclear motion is best described by
classical Newton dynamics instead of the time-dependent Schr{\"o}dinger equation
is substantiated. In principle a single experiment should bring clarity. Caution is
however necessary, as temperature dependent effects must be eliminated when trying to
determine the existence of a zero-point energy.
\end{abstract}

\pacs{03,31}
\maketitle

Recently \cite{Frank2014}, we argued that nuclear motion is best described classically instead of using the time-dependent Schr{\"o}dinger equation. A victory for Car and Parrinello who chose to develop a code which treats the electronic cloud quantum mechanically (using the density functional theory (DFT) approximation) and the motion of the nuclei classically \cite{Car1985}. Car-Parrinello molecular dynamics (CPMD) calculations and, more general, ab-initio molecular dynamics (AIMD) calculations (not necessarily using DFT) were always considered to be an approximation to a proper quantum mechanical treatment of the nuclei which would use the time-dependent Schr{\"o}dinger equation. To get things sorted: within AIMD there are different ways to describe the propagation of the electrons, in particular, CPMD and BOMD (Born-Oppenheimer MD). For the present consideration this difference is irrelevant. We are focusing in this paper on the motion of the nuclei, not touching the motion of the electrons any. Also it should be mentioned that the whole consideration ignores relativity; for the nuclei relativity can be used just as well, the numerical difference is negligible since the nuclei move slowly, approximately with the speed of sound, far from the speed of light. The term 'classical' is used as difference to quantum mechanics, not to relativity.

Why take the simulation of chemical reactions so important in this context? Because the results of such simulations were perfectly unpredictable. Schr{\"o}dinger developed his equation for the hydrogen atom with one electron only. Pauli added spin and the Pauli principle to describe systems with more than one electron and this all a century ago when it was impossible to check anything numerically with computers. A century later one is used to interpreting orbitals and their imagined motion during a reaction. A lot is guessing with chemical intuition. Our movies frequently confirm this chemical intuition, sometimes no. Sometimes we learn that a multi-step mechanism is a trivial explanation for a difficult to understand reaction pattern.

Nevertheless, that we succeeded to describe a finite number of experiments might be accident and is only a moderately strong argument. But there must be lots of evidence, right? There are molecular vibrations and the spectra are well known.
Classical motion means there is no zero-point energy. The well-known  $\frac{1}{2} h\nu$ would be missing, independent if one looks at a quadratic potential, or a Morse potential, or if one computes the potential explicitly. Should one not see from the spectra that a quantum mechanical description is clearly superior?

Actually, by fourier transform of CPMD trajectories, it is possible too, to obtain a quantitatively reasonable description of vibrational spectra \cite{Silvestrelli1997}, a convenient approach which can easily also be applied to large molecular systems. (For more quantitative results one has to use lower electron masses and time steps.) The zero-point energy is not directly seen in the spectra, since, like in experiment, only energy differences are measured.



Have a look at some of our movies of highly reactive systems \cite{Frankpage} to get a realistic picture of nuclear motion. Even if our pictures are still learning how to run, one gets a general impression. If someone would describe nuclear motion one day more accurately, optically the 3D motion will more or less look like in our movies now, not so much will change. At normal temperatures (T $<<$ 10000 K) molecules are most of the time rather rigid objects. All bonds are slightely vibrating and their velocities obey a Maxwell-Boltzmann distribution. A bond breaking is a rare event. This can be changed in various ways, for example by applying an external electrical field in a CPMD simulation to a stable system. First, it is possible to get resonance at certain frequencies. The molecules take up energy whenever the frequency applied enhances a certain vibration or rotation. Overtones are possible. These sinusoidal motions lead to peaks upon Fourier transformation. Then one can also make, as we checked for HCl, a single bond dissociate, if one irradiates the molecule with an electrical field with linearly decreasing frequency (Fig. 1). A linearly changing field allows the molecule to stay in resonance when climbing up the Morse potential -- remind the Birge-Sponer plot. If a constant frequency is applied instead, the molecule gets out of resonance after some time and relaxes again.

\begin{figure}
\includegraphics[height=6cm]{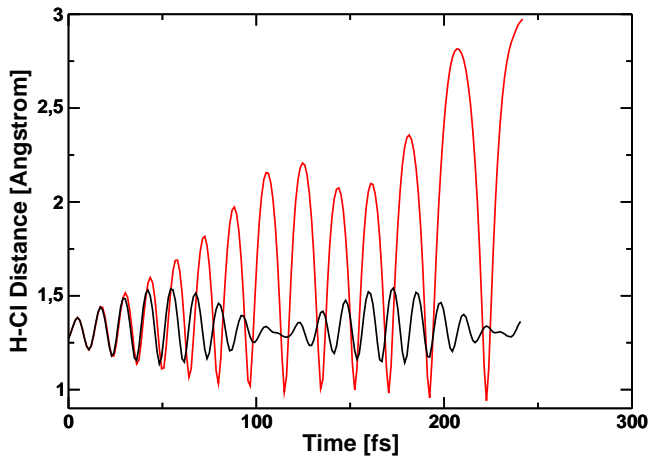}
\label{Fig1}
\caption{
Example for the classical treatment of the nuclei with CPMD: HCl dissociates if a field with a linearly decreasing frequency is applied (red line). It relaxes again if instead the frequency of the external field is constant (black line). Our usual methodology was applied (CPMD, LSD, BLYP, Troullier-Martins pseudopotentials, see for example \cite{Alznauer2013}).}
\end{figure}

But is there not an isotope effect in nuclear motion?

Let us have a look. When are isotope effects observed? What experiment gives a reliable answer? I recommended \cite{Frank2014} to directly measure the reaction enthalpy $\Delta\Delta H_r$ for the reactions $3 H_2 + N_2 \to 2 NH_3$ and $3 D_2 + N_2 \to 2 NH_3$ using calorimetry. The ammoniak formation changes the bond situation strongly from a triple to single bonds which makes the zero-point energy effect relatively high (3 kcal/mol) compared to other reactions. The water formation from the elements is another example where calorimetry could yield a result that is beyond doubt.
The values given in data bases \cite{NIST2013,Chase1985} are computed under the assumption that there must be a zero-point energy in nuclear motion. The conviction that objects below a certain size must be treated quantum mechanically may have been strong enough to affect the believe in calorimetry experiments. Now let me argue that it is not the size alone which determines the differential equation to be used for the description of a certain object. It may have been a consequence of a simplified believe in quantum theory --- the Schr{\"o}dinger equation must be used everywhere --- that chemists tried with all means to prove the zero-point energy via an isotope effect. They searched then for a kinetic isotope effect instead of the thermodynamic isotope effect. Indeed, this kinetic isotope effect of deuterisation on the velocity of chemical reactions is easily observed. It turns out however that also Newton dynamics with the equation ${\bf F_i} = m  {\bf a_i}$ depends on the mass and for a given potential, $- \frac{\partial V}{\partial \bf{x_i}} = {\bf F_i}$, yields lower reaction velocities if the mass of an object is higher.

The same accounts to the observation of osmotic pressure for deuterated compounds which also sometimes is termed a thermodynamic isotope effect. This experiment has nothing to do with mass, no matter what one assumes about nuclear motion. The osmotic pressure results from probability. Two somehow different liquids brought into contact will mix, no matter by what the two liquids differ. This experiment does not help to discriminate classical or quantum mechanical motion of the nuclei.

To conclude, the experimental evidence is rather uncertain. When now is the Schr{\"o}dinger equation to be applied?

Answer: To bound electrons, for which Schr{\"o}dinger developed his equation. 

How do we know?

Not at all, for the moment we have to compare to experiment. The one and only equation describing everything is still missing, and in particular its decomposition into separate differential equations for the nuclei, the electronic cloud and the motion of the complete system. So, we look at electronic clouds as part of an atom or a molecule. We separated it from the nucleus. No point in assuming that this latter object which is so different, obeys the same differential equation. Think of the eye of a hurricane or the center of a whirl in a bathtub. How would one describe such an object, the center of a whirl or vortex? Computational fluid dynamics calculations carried out on simple systems and using the language of quantum chemists \cite{Levine2000} could help to get an idea. As a first guess, the motion of such an object may be classical with some mass resulting from its extension, and the inner structure of its nucleus or eye might be best described by a constant density. That indeed reminds to existing models in nuclear physics. As long as we are looking at chemistry only, we can continue to treat the nucleus as a moving point charge.

What happened to Schr{\"o}dinger’s cat when doing so?

It died, in its box, perfectly deterministically, perfectly obeying all relevant differential equations, and there are many different techniques to find out when exactly it died. Noise, heat, electromagnetic radiation in some range, there is always something the box will easily let through. That is true for any physically existing box in our universe. A linear combination between dead and alive, as long as not measured: No, not in this physical universe.

Obviously we lose fancy philosophical concepts around quantum mechanics, but knowing if ammoniak and water formation show a zero point energy 'at zero Kelvin' or not would give us a fix point from which one could think about the differential equations employed in a quantum field theory again. This seems indeed close and and it is without doubt that quantum field theory is the basis of it all in the end.
To say something about the electrons as an outlook: In BOMD the electrons follow immediately the nuclei, the electronic wavefunction is fully converged to the new positions of the nuclei in every step. In CPMD the electrons have a fictitious mass $\mu$ and follow like in a Newton dynamics with some delay. Remarkably the resulting equations have some similarity with the celebrated Klein-Gordon equation which is a relativistic quantum field equation.

Car-Parrinello equations:
\begin{equation}
\left[\mu \frac{\partial^2}{\partial t^2} - \frac{\hbar^2}{2m}\left(\frac{\partial^2}{\partial x^2} + \frac{\partial^2}{\partial y^2} + \frac{\partial^2}{\partial z^2}\right) + v_{eff} \right] \psi_i =  \sum_j \epsilon_{ij}  \psi_j
\end{equation}

Klein-Gordon equation:
\begin{equation}
\left[\frac{\hbar^2}{m c^2}\frac{\partial^2}{\partial t^2} - \frac{\hbar^2}{m}\left(\frac{\partial^2}{\partial x^2} + \frac{\partial^2}{\partial y^2} + \frac{\partial^2}{\partial z^2}\right) + m c^2 \right] \Psi = 0
\end{equation}

The first set of equations is well tested by all the Car-Parrinello molecular dynamics simulation runs. Obtained are the perfectly real orbitals $\psi_i$ which depict most concrete a chemical structure or reaction. It is easy to interpret the change of these clouds. Some chemical questions are easier to interpret in terms of localized orbitals, but also these localized orbitals extend over an angstrom. Localization of orbitals down to point charges changes the physics, but can help to illustrate phenomena like charge transfer. It is these perfectly deterministicly moving clouds which determine a molecular structure or a chemical reaction, every one of them determined by the effective potential v$_{eff}$ formed by all other particles in the system. This should remove the quantum mechanical observation problem. 

Numerically there is a lot of problems unsolved. From the comparison of the two equations above it is evident that the fictitious electron masses in CPMD are way too high. Hence it will not be easy to directly simulate electronic excitations. That also accounts to attempts with the time-dependent Schr{\"o}dinger equation while with perturbation theory this equation is successfully used in describing electronic spectra. So, at the moment there a lots of open questions, but we have a very interesting system of equations to investigate.

\end{document}